\documentclass[a4paper]{jpconf}
\usepackage{graphicx}
\usepackage{hyperref}
\newcommand{\unit}[1]{~\mathrm{#1}}

\newcommand{\zu}[1]{Fig.~{\ref{#1}}}

\def\rtHz{\sqrt{Hz}}
\begin{document}
\title{The status of KAGRA underground cryogenic gravitational wave telescope}
\author{T.~Akutsu$^{1}$, M.~Ando$^{2}$, A.~Araya$^{3}$, N.~Aritomi$^{2}$, H.~Asada$^{4}$, Y.~Aso$^{5}$, S.~Atsuta$^{6}$, K.~Awai$^{7}$, M.~A.~Barton$^{1}$, K.~Cannon$^{8}$
% ...
(KAGRA Collaboration)}
\address{$^{1}$ National Astronomical Observatory of Japan, Mitaka, Tokyo 181-8588, Japan}
\address{$^{2}$ Department of Physics, University of Tokyo, Bunkyo, Tokyo 113-0033, Japan}
\address{$^{3}$ Earthquake Research Institute, University of Tokyo, Bunkyo, Tokyo 113-0032, Japan}
\address{$^{4}$ Department of Advanced Physics, Hirosaki University, Hirosaki, Aomori 036-8561, Japan}
\address{$^{5}$ National Astronomical Observatory of Japan, Hida, Gifu 506-1205, Japan}
\address{$^{6}$ Department of Physics, Tokyo Institute of Technology, Meguro, Tokyo 152-8551, Japan}
\address{$^{7}$ Institute for Cosmic Ray Research, University of Tokyo, Hida, Gifu 506-1205, Japan}
\address{$^{8}$ Research Center for the Early Universe, University of Tokyo, Bunkyo, Tokyo 113-0033, Japan}
% ...
\ead{michimura@granite.phys.s.u-tokyo.ac.jp}

\begin{abstract}
KAGRA is a 3-km interferometric gravitational wave telescope located in the Kamioka mine in Japan. It is the first km-class gravitational wave telescope constructed underground to reduce seismic noise, and the first km-class telescope to use cryogenic cooling of test masses to reduce thermal noise. The construction of the infrastructure to house the interferometer in the tunnel, and the initial phase operation of the interferometer with a simple 3-km Michelson configuration have been completed. The first cryogenic operation is expected in 2018, and the observing runs with a full interferometer are expected in 2020s. The basic interferometer configuration and the current status of KAGRA are described.
\end{abstract}

\section{Introduction}
The direct detections of gravitational waves by Advanced LIGO in 2015 opened a completely new frontier in astronomy~\cite{GW150914}. In 2017, Advanced Virgo~\cite{AdVirgo} started simultaneous observing run with Advanced LIGO to extend the global network of advanced gravitational wave telescopes, and clearly demonstrated that better sky localization is possible with a network of three detectors~\cite{GW170814}. Coincident detection of gravitational-waves at multiple sites further helps sky localization and parameter estimation of the source~\cite{ObservationScenario}.

KAGRA is another ground-based advanced interferometric gravitational wave telescope, which aims to further enhance the field of gravitational wave astronomy~\cite{SomiyaKAGRA,AsoKAGRA}. Although the basic interferometer configuration is very similar to other telescopes, KAGRA uses two distinct strategies. Advanced LIGO and Advanced Virgo are constructed on the surface of the Earth, whereas KAGRA is constructed at an underground site in the Kamioka mine in Japan to reduce seismic noise, gravity gradient noise and other environmental fluctuations such as temperature and humidity~\cite{UchiyamaTunnel}. Advanced LIGO and Advanced Virgo use room temperature fused silica test masses, whereas KAGRA uses sapphire test masses at around 20~K to reduce thermal noise~\cite{HiroseSapphire,HiroseCoating}. Some concepts of future gravitational wave telescopes~\cite{ET,CE} plan to incorporate underground construction and cryogenic cooling of test masses, and KAGRA is expected to pioneer technologies for the future.

The construction of the initial phase facility including the tunnel, vacuum system, cryostats, and clean booths was completed in November 2015. In March and April 2016, we have performed the very first operation of a simplified 3-km Michelson interferometer at room temperature. Our next step is to operate the cryogenic interferometer, and various installation works and tests are underway. In this article, we briefly describe the interferometer configuration and the current status of the KAGRA project.

\section{Interferometer Configuration}
KAGRA interferometer is a resonant sideband extraction (RSE) interferometer with 3-km Fabry-P{\'e}rot arm cavities formed by sapphire input test masses (ITMs) and end test masses (ETMs) at cryogenic temperatures (see \zu{fig:Config}). Other mirrors such as a beam splitter (BS), power recyling mirrors (PRM, PR2, PR3) and signal recycling mirrors (SRM, SR2, SR3) are room temperature fused silica mirrors. Power recycling cavity enhances the effective input power, and signal recycling cavity broaden the detector bandwidth by changing the spectral shape of the quantum noise. The interferometer is equipped with input and output mode cleaners (IMC and OMC) to reject higher-order spatial modes and unwanted frequency sidebands of the input and output beams. IMC is a triangular ring cavity formed by three suspended mirrors, and has a round-trip length of 53.3~m and finesse of 540. OMC is a bow-tie cavity formed by four mirrors monolithically fixed on a base plate, and has a round-trip length of 1.5~m and finesse of 780~\cite{KAGRAOMC}. The gravitational wave signal is readout from the DC power of the OMC transmitted beam.

\begin{figure}[t]
	\begin{center}
		\includegraphics[width=14cm]{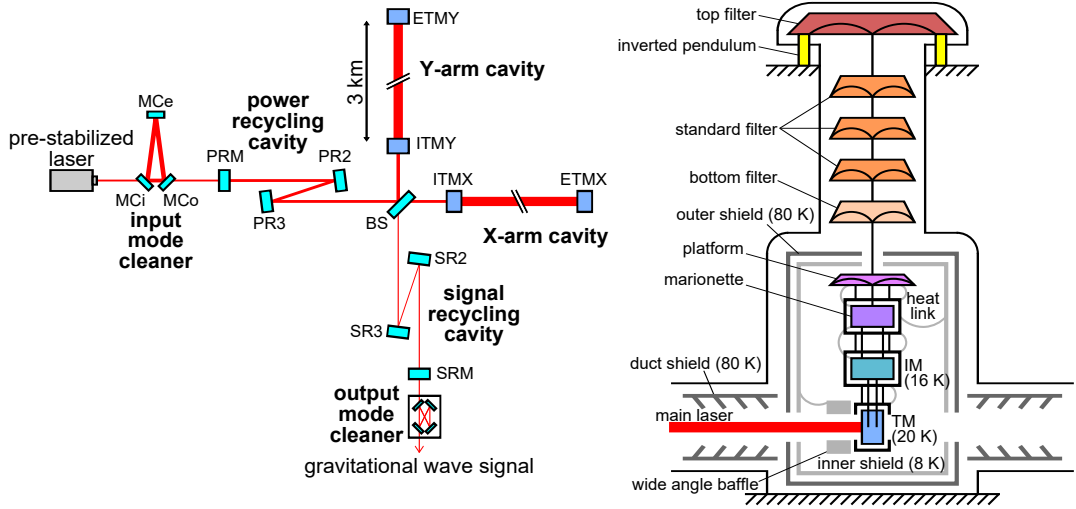}
	\caption{Schematic of the KAGRA interferometer (left) and cryogenic suspension system for sapphire test masses (right). Marionette, intermediate mass (IM) and test mass (TM) are surrounded by respective recoil masses for position and alignment control.}
	\label{fig:Config}
	\end{center}
\end{figure}

We use 180-W continuous-wave laser source at a wavelength of 1064~nm. The laser frequency is pre-stabilized with respect to the IMC length, and the laser intensity is stabilized by monitoring the IMC transmitted power. The laser frequency is ultimately stabilized with respect to the averaged length of the 3-km arm cavities. The designed power recycling gain is 10 and the arm cavity finesse is 1530. With input power to PRM of 78~W, the power inside the arm cavities reaches 400~kW. The intra-cavity power is almost as half as that of Advanced LIGO or Advanced Virgo design for cryogenic cooling of test masses.

ITMs and ETMs are suspended by an eight-stage pendulum suspended from a top geometric anti-spring filter on an inverted pendulum table for low frequency vertical and horizontal vibration isolation~\cite{TypeA} (see \zu{fig:Config}). The last four stages of the pendulum is cooled down to cryogenic temperatures and called a cryogenic payload~\cite{Cryopayload}. The heat generated by the main intra-cavity beam is extracted via sapphire fibers which suspend the test mass from the intermediate mass, and via high purity 6N aluminum heat links. The thermal conductivity of the sapphire fiber is measured to be $\kappa \simeq 7.98 \times (T/1 \unit{K})^{2.2} \unit{W K^{-1} m^{-1}}$ at cryogenic temperatures, which is high enough to cool down the mirror to 20~K~\cite{SapphireFiber}. The cryogenic payload is surrounded by 8~K inner shields and 80~K outer shields, which are cooled down by low vibration pulse-tube cryocoolers~\cite{SakakibaraCryo}. The test masses also have different kinds of baffles for absorbing stray light inside the cavity, and duct shields at both sides for absorbing thermal radiation from ducts at room temperature~\cite{SakakibaraShield,AkutsuBaffle}.

Room temperature mirrors are suspended from simpler vibration isolation systems. BS and signal recycling mirrors are each suspended by a four-stage pendulum from a top geometric anti-spring filter on an inverted pendulum table. Power recyling mirrors are each suspended by similar system, but by a triple-pendulum and has no inverted pendulum. IMC mirrors are suspended by a double pendulum fixed on vacuum compatible vibration isolation stacks. All the mirrors are installed in the vacuum chambers at $10^{-7} \unit{Pa}$ to mitigate noise from residual gas. For details of the interferometer configuration and room temperature suspension systems, see Refs.~\cite{SomiyaKAGRA,AsoKAGRA} and Refs.~\cite{FabianRSI,KAGRAActuator}, respectively.

Along the KAGRA interferometer, environmental sensors such as thermo-hygrometers, barometers, seismometers, magnetometers, and microphones are placed to monitor and characterize environmental transient noises~\cite{SasakiEnv}. One of the most significant sensors is a 1.5-km laser strainmeter, which is placed along the X-arm. This strainmeter is fixed to the ground to monitor low-frequency ground motion, and has been operating since August 2016~\cite{ArayaGIF}.

\section{iKAGRA Operation}
\begin{figure}
	\begin{center}
\begin{minipage}[b]{0.45\textwidth}
   \begin{center}
   \includegraphics[width=7cm]{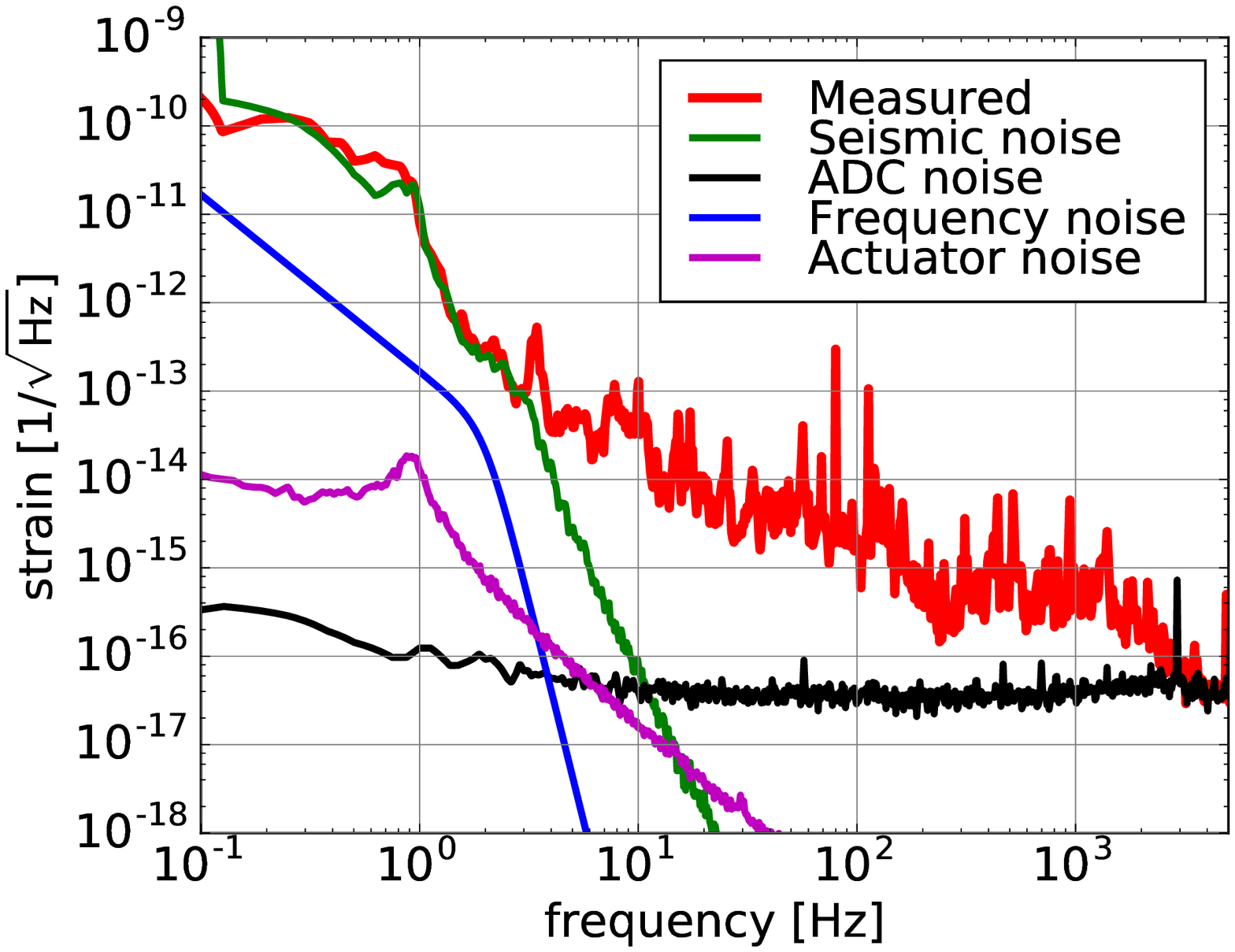} \\
   \end{center}
\end{minipage}   
\begin{minipage}[b]{0.54\textwidth}
   \begin{center}
   \includegraphics[width=8cm]{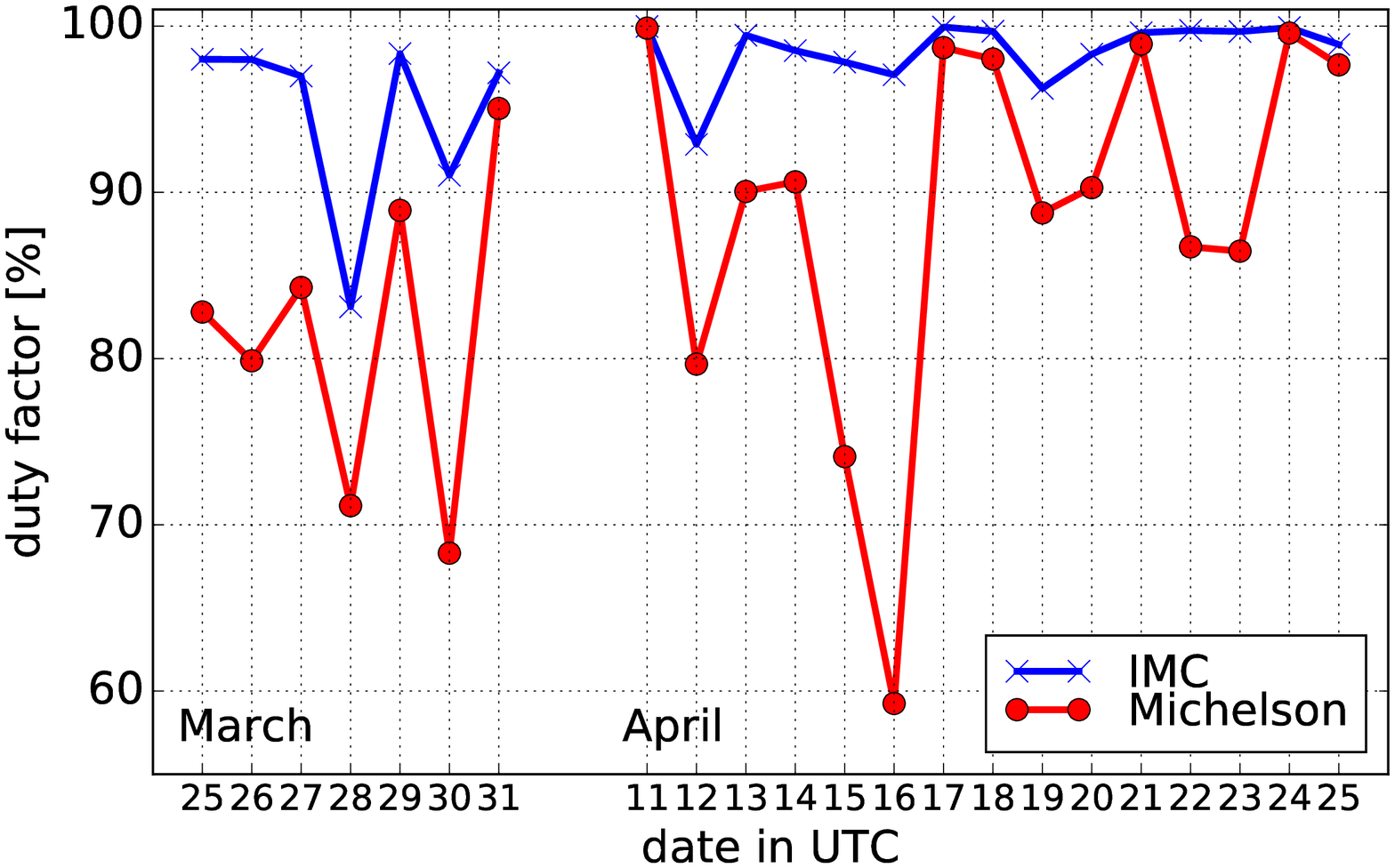} \\
   \end{center}
\end{minipage}
	\caption{Strain sensitivity (left) and daily duty factor (right) of iKAGRA. The sensitivity plotted here does not represent the best sensitivity. Peaks at 80~Hz and 113~Hz are from calibration line injections. Significant degradation in the duty factor on April 16 is due to an earthquake with a magnitute 7.3 which hit Kumamoto at 16:25 UTC.} \label{fig:iKAGRAResult}
	\end{center}
\end{figure}

The KAGRA project is split into two stages, initial phase (iKAGRA) and baseline phase (bKAGRA). iKAGRA was aimed to operate the 3-km underground interferometer for the first time, and bKAGRA aims to operate the full cryogenic RSE interferometer for observing runs.

In March and April 2016, we have operated the 3-km Michelson interferometer for the first time, with significantly simplified mirror suspension systems. ETMs were room temperature fused silica mirrors suspended by a double pendulum, and only PR2, PR3 and BS were installed other than ETMs and IMC mirrors. All the mirrors were installed inside the vacuum chambers at air pressure, except for the IMC mirrors at 200~Pa. We used a 2-W laser source and input power to the Michelson interferometer was about 200~mW. The interferometer control was done by extracting the error signals from the Michelson reflection port and feeding back the control signal to actuators on ETMX and ETMY differentially. With this iKAGRA operation, we have tested the basic performance of the digital real-time interferometer control system and data acquisition, transfer and analysis pipelines.

The iKAGRA operation was done in two periods, from March 25 to 31 and from April 11 to 25. Between the March run and the April run, we had a little commissioning break to improve the sensitivity and the stability of the interferometer. The strain sensitivity in March and April was $3 \times 10^{-15} \unit{/\rtHz}$ and $6 \times 10^{-16} \unit{/\rtHz}$ at 100~Hz, respectively. The duty cycle, the ratio of locked period to whole run period, in March and April was 80\% and 89\%, respectively.
These improvements were done mainly by changing the lock scheme from mid-fringe lock to dark-fringe lock using frontal modulation technique, and by improving the actuator balancing of the ETMs.

The strain sensitivity was limited by seismic noise below 3~Hz and analog-to-digital converter noise above 3~kHz (see \zu{fig:iKAGRAResult}). The limiting noise source in mid-frequencies is not completely understood, but acoustic noise coupling was proved to be very large because mirrors were installed at air pressure. After April run, some of the fans of the clean booths were turned off to show that the sensitivity can be improved by reducing acoustic noise. The best sensitivity achieved after the test run was $2 \times 10^{-16} \unit{/\rtHz}$ at 100~Hz.

The acquired gravitational wave signal and environmental monitor signal data were transferred to the data centers at Institute for Cosmic Ray Research in Kashiwa and Osaka City University with the latency of about 2.5 to 3 seconds. The data transfer rate was about 20~GB/hour, and the total amount of data was about 7.5~TB. Searches for bursts, compact binary coalescence, and continuous gravitational waves using the acquired data were done. Details of the iKAGRA construction and operation, and results of the data management and data analysis will be discussed elsewhere~\cite{iKAGRAConstruction}.
\section{Outlook}
After the completion of iKAGRA operation, we have quickly transitioned to the bKAGRA phase. bKAGRA is split into 3 phases by project milestones, the first operation of 3-km cryogenic Michelson interferometer (Phase 1), the first operation of full cryogenic RSE interferometer (Phase 2), and the start of observing runs (Phase 3).

At the time of writing, we are currently installing the mirrors for bKAGRA Phase 1. PRM, PR2, PR3 and BS suspension systems were already installed. Installation of the room temperature part of the ETMY suspension was completed, test assembly of the cryogenic part was also completed, and cooling test with a full suspension chain showed that the test mass can be cooled down to 12~K in 23~days, without any heatload from the laser beam. Assembly of ETMX suspension is also underway, and once completed, we will operate the 3-km cryogenic Michelson interferometer by March 2018. The signal recycling mirrors and ITMs will be installed by the end of 2018. After the installation of all the mirrors, we will start the commissioning of each arm and central dual-recycled Michelson iterferometer step by step. We are expecting the first full operation one year after the completion of the installation work, and observing runs are expected to start in early 2020s.

The sky position of many gravitational wave signals can be determined to below $10 \unit{deg^2}$ with coincident detection by Advanced LIGO, Advanced Virgo and KAGRA~\cite{ObservationScenario}. Detection by multiple detectors with different orientations also resolve the distance and binary inclination degeneracy from polarization measurements. Contribution of KAGRA will be significant for multi-messenger astronomy and precision gravitational wave astronomy.

\section{Summary}
The major construction of KAGRA has been completed and the initial phase operation was performed. With the initial phase operation of 3-km Michelson interferometer, we have tested the basic system from interferometer controls to data analysis pipelines as a gravitational wave telescope. We are now in the installation phase for the first cryogenic operation, and the observing runs with the full interferometer are expected in early 2020s. KAGRA joining the global network of advanced gravitational wave telescopes is important for sky localization and parameter estimation of the source. With underground and cryogenic technologies, KAGRA also paves the way to future gravitational wave telescopes.

\ack
The KAGRA project is supported by MEXT, JSPS Leading-edge Research Infrastructure Program, JSPS Grant-in-Aid for Specially Promoted Research 26000005, MEXT Grant-in-Aid for Scientific Research on Innovative Areas 24103005, JSPS Core-to-Core Program, A. Advanced Research Networks, the joint research program of the Institute for Cosmic Ray Research, University of Tokyo, National Research Foundation (NRF) and Computing Infrastructure Project of KISTI-GSDC in Korea, the LIGO project, and the Virgo project.

\section*{References}
\bibliographystyle{iopart-num}
\providecommand{\newblock}{}

\end{document}